# Static Analyzers and Potential Future Research Directions for Scala: An Overview

Eljose E Sajan[1], Yunpeng Zhang[2*], Liang-Chieh Cheng[3]

[1]College of Natural Science and Mathematics, University of Houston, eesajan@central.uh.edu
[2]College of Technology, University of Houston, yzhan226@central.uh.edu
[3]College of Technology, University of Houston, lcheng6@Central.uh.edu

**Abstract**—Static analyzers are tool-sets which are proving to be indispensable to modern programmers. These enable the programmers to detect possible errors and security defects present in the current code base within the implementation phase of the development cycle; rather than relying on a standalone testing phase. Static analyzers typically highlight possible defects within the 'static' source code and thus does not require the source code to be compiled or executed. The Scala programming language has been gaining wider adoption across various industries in recent years. With such a wide adoption of tools of this nature, this paper presents an overview on the static analysis tools available, both commercial and open-source, for the Scala programming language. This paper discusses in detail about the types of defects that each of these tools can detect, limitations of these tools and also provide potential research direction that can improve the current state of static analyzers for the Scala programming language.

## I. INTRODUCTION

Software development is still a human initiated task that requires extensive time and effort to plan and implement. Programmers and developers work with very short deadlines and are most concerned about delivering functionalities rather than having application security as the primary objective. Hence most software designs forgoe code testing and verification during the development stages [1]. Software defects, or bugs, can cost companies significant amounts of money, especially when they lead to software failure [2]. Based on the estimates made by the National Institute of Standards and Technology (NIST), the US economy loses $60 billion annually in costs associated with developing and distributing patches that fix software faults and vulnerabilities, as well as costs from lost productivity due to computer malware and other problems caused by software faults [3]. Hence detecting defects early on during the development phase is critical for developing secure and functional software that could potentially save the developers time, resources and money in the lifetime of the software. Static analysis tools could provide a means for rectifying this problem, where code can be analyzed without having to run the code, helping ensure higher quality software throughout the development process.

Static code analysis of a program generally involves an automated tool that takes as input the source code (or object code/ byte code in some cases) of the program, examines this code without compiling or executing it, and yields results that usually indicates the location and types of defects that the given code can potentially contain. Static analyzers accomplish this by checking the code structure, the sequences of statements, and how variable values are processed throughout the different program segments. The main advantage of static analysis is that every line of the source code is analyzed, which differs from dynamic analysis where certain portions of code could only be executed under some specific conditions, which in most likelihood could never be met during the analysis phase and would require actual code compilation or execution. A typical static analysis process starts by representing the analyzed source code to some abstract representation such as an Abstract Syntax Tree (AST) or other representations based on the purpose of analysis. Those abstract models provide a simplified interface for supporting upper level client analyses. Other information, such as the values of variables at different statements of the CFG can also be collected to allow the static analysis to support more in-depth verification, e.g., through data-flow analysis. This paper mainly focus on static analysis technologies in Scala

## II. SCALA LANGUAGE

Scala is a general purpose programming language that combines both object-oriented programming and a functional programming paradigm. Some of its design decisions were directly influenced to overcome some of the limitations of Java [4]. The current version of Scala 2.12 provides 3 runtime environments, including the default and widely used Java Virtual Machine (JVM), JavaScript runtime targeting front-end web development, and a Native runtime environment for Scala built using LLVM principles (which is still in beta). Since the most widely used runtime is the JVM, this is what will be considered in this paper moving forward. Since Scala can use the JVM as a runtime, this allows Scala programs to import and use the vast external libraries that are available in Java. Scala was designed with a pure object-oriented model, i.e. a uniform object model, which conceptualizes every value as an object and any operation to be performed as a function call [5]. The functional programming paradigm implies that functions are treated as first-class values. As every value is treated as an object, then this means that every function in Scala is treated as a first-class object. Scala also has other functional programming features, including lazy evaluation, pattern-matching (which can model the algebraic types used in other functional languages) and a powerful type-inference system.

## III. CURRENT STATE OF STATIC ANALYSIS TOOLS AND RESEARCH TARGETING SCALA

The Scala programming language has been gradually rising in popularity in the development sphere. Figure 1, represents the overall growth of Scala popularity based on the TIOBE index. As of the date of this document Scala has a rank of 28, and it had the highest rank in 2018 with a rank of 20. Scala shows a positive trend in popularity, but our research has found that there is a significant lack of Static Analysis tools for Scala.

### A. CheckMarx

We find that Checkmark's CxSAST is the only commercially available product that supports static code analysis in Scala. According to the product page [6], CxSAST can detect 26 Scala specific vulnerabilities, including code injection, SQL injection and Absolute Path Traversals, to name a few. It is to be noted that CxSAST is not a language or platform specific product. It is designed to be platform and language agnostic, meaning that a single instance of this product can be used to perform static analysis on source codes written in different languages.

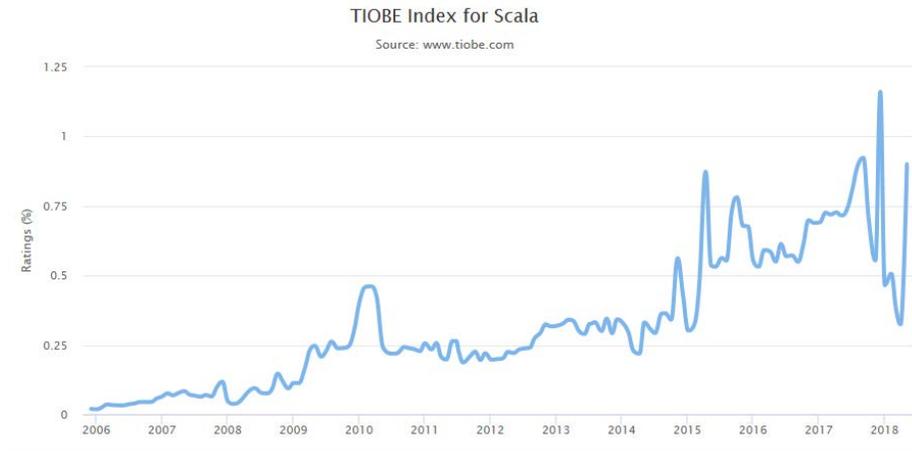

Figure 1. The growth of popularity of Scala programming language
Courtesy TIOBE; www.tiobe.com/tiobe-index/scala/

One of the main advantages or more accurate features that led to the large adoption rate of this specific tool by the developer community are:

i. Platform/compiler agnostic design: In a modern development environment, a software product is usually composed of numerous programming languages. Hence it would become impractical for the development team to use different source code checkers for different segments of the product.

ii. CxSAST is available as an IDE plugin: Even though the CheckMarx architecture performs the actual code analysis in a dedicated Server, the results of the analysis are presented back in IDE plugin format [7].

As stated before, the CxSAST tool for Scala can detect only 26 Scala specific vulnerabilities. Thirteen of these vulnerabilities are specified on the company website, but the remaining ones, upon initial search, are not disclosed. This lack of information available about the checks performed was present for other Language implementation of CheckMarx, as well as other main stream Static analysis tools [8].

B. *SonarScala*

SonarScala is an open-source static code analyzer developed by SonarSource, specifically designed to detect vulnerabilities and code bugs in Scala. Similar to CheckMarx, the SonarSource static tool is designed to be Language agnostic, at least to some degree, in its internal architecture. This allows support for multiple programming languages, almost 25 languages as of the writing of this survey.

SonarScala provides 41 rules, detailing the different vulnerabilities that it can detect. By SonarSource definitions, the issue types into which different vulnerabilities are categorized are as follows [9]:

i. Code Smells: Maintainability issues are termed as code smells by SonarSource. These include minor issues which are not critical in nature by themselves, but issues that can reduce efficiency of code development. These include concepts, such as modularity, understandability, changeability, testability and reusability.

ii. Bugs: Reliability issues are termed as Bugs by SonarSource. This category of issues groups everything that has to do with operational risks or unexpected behavior at runtime.

iii. Vulnerabilities: Security issues are termed as Vulnerabilities by SonarSource. This category of issues groups everything that has to do with a program having flaws that can be exploited to make it behave differently from what it was designed for.

Table 1 represents the bugs and vulnerabilities that are detectable by SonarScala; it does not consist of entries of type code smells since it mainly focuses on the lack of best coding practices rather than critical defects. From this it is obvious to the readers that the number of high-risk defects detectable by SonarScala is an extremely limited 8 for the whole language. This is starkly less than the detection rate of SonarJava, another product by SonarSource for Java, which has almost 538 rules, of which 236 are targeted to detect high risk defects [10]. This disproportion in the number of high-risk defects that are detectable by these commercially available static analyzers could be attributed to the fact that Java is a more popular language than Scala and has a much higher adoption rate in the development community. Nevertheless, it is quite clear that there is a large limitation on the number of defects, at least those of greater consequence, which can be detected by SonarScala.

TABLE 1. Scala specific rules of SonarScala and corresponding issue types

| Rule | Issue Type |
|---|---|
| Credentials should not be hard coded | Vulnerability |
| Related "if"/"else if" statements and "case" in a "match" should not have the same condition | Bug |
| Identical expressions should not be used on both sides of a binary operator | Bug |
| Jump statements should not be followed by dead code | Bug |
| Variables should not be self-assigned | Bug |
| Useless "if(true) {...}" and "if(false){...}" blocks should be removed | Bug |
| Using hard-coded IP addresses is security insensitive | Security hotspot |

The SonarScala static analyzer is built using the open-source slang framework. This framework allows creation of language agnostic AST, thereby allowing them to quickly build simple syntax-based rules for their platform [11]. After this a language specific parser is used to create an AST that adheres to the language grammar. SonarScala uses the ScalaMeta parser library as the Scala specific parser.

The SonarScala static analyzer is built using the open-source slang framework. This framework is designed by SonarSource and allows creation of language agnostic AST, thereby allowing them to quickly build simple syntax-based rules for their platform [11]. After this a language specific parser is used to create an AST that adheres to the language grammar. SonarScala uses the ScalaMeta parser library as the Scala specific parser.

*C. ScalaStyle*

ScalaStyle is one of the more famous open-source static analysis tools available for Scala; used as part of the grading framework by Martin Odersky for his course [12].

It should be noted that, unlike the previous tools discussed, ScalaStyle was specifically designed for style checking and to check if the given source code complies with coding conventions. ScalaStyle can be incorporated into one development workflow in a number of ways: use a standalone command line tool, run ScalaStyle as a compiler plugin (SBT, Maven and Gradle are supported) or integrate it in an IDE (IntelliJ or Eclipse). By default, ScalaStyle has 69 rules specified. These rules range from code style to best practices, such as checks that null is not used, checks so that not too many types are declared in a file and checks that ensure Java @Override is not used, to name a few.

Although ScalaStyle does provide developers the option to implement custom rules, it is up for debate if one can configure rules that could detect high risk vulnerabilities since by design, ScalaStyle was intended to detect only styling defects.

*D. WartRemover*

This is an open-source Scala code analyzing tool that is specifically designed for detecting a wide range of code issues or "warts". It comes with support for 37 predefined warts [13], but a prominent feature of WartRemover is its custom wart extensibility. This allows developers to write custom check rules to detect specific defects in the code. Unlike ScalaStyle, which is designed to detect styling specific defects in code, WartRemover is more general and can detect defects that could potentially lead to compile time errors.

*E. Scapegoat*

Scapegoat is another open source tool which has similar design principles to ScalaStyle. But unlike ScalaStyle, which focuses primarily on styling and best code standards, Scapegoat provides static detection capabilities for defects that have a higher severity to program health. Scapegoat is developed as a Scala compiler plugin, which can then be used inside a project's build tool. It can be integrated into the project workflow via SBT, Maven or Gradle [14]. A rule defining the defect check is termed as an "inspection" in Scapegoat vocabulary. As of the writing of this report, Scapegoat includes 117 predefined inspections. Of these 117 inspections, 21 are of the type "Error", denoting the highest severity that would lead to potential problems in the program's lifecycle. The predefined inspections also consist of 47 inspections of default level "info" which encompasses checks on styling and best code practices. The final group of inspections are called "warnings" and Scapegoat comes with 49 predefined warnings. Warnings have a medium severity level and denote defects in the given code that will lead to compile time or run time errors in most cases. Table 2 provides a sample of the inspections provided by Scapegoat.

IV. CONCLUSIONS AND FUTURE WORK

From the previous section it is apparent that most publicly available Scala static analyzers, whether commercial or open-source, is extremely limited in the scope and nature of defects that can be detected. In a language like Scala, with features such as type-inference, pattern- matching [15], designing a universal static analyzer that can detect a large case of errors is extremely difficult. Nevertheless, the number and type of serious errors that are being detected by the surveyed tools are extremely limited when compared to another similar language, Java. Checkmark's CxSAST tool provides the most coverage for defects of potentially higher severity, partly due to the fact that its design principle is based on delivering a minimum coverage of all the OWASP 10 and SANS 25 vulnerabilities. But due to the proprietary nature of its tools and techniques, these claims could not be verified with certainty. Although SonarScala is Open-source, the tool only covers 8 defects that could be of significance. The rest of the detection rules for SonarScala as well as most rules for ScalaStyle focus mainly on code styling and best practice suggestions. SonarScala and WartRemover provide capabilities for an end user to develop their own custom rules to detect specific defects, if this target defect is not provided within their included rules. Below a list of certain known vulnerabilities and defects in Scala are provided and we discuss whether the current tools and techniques available can detect these. To conclude, we propose potential future research directions.

TABLE 2. Sample of Inspection rules defined for Array operations

| Name | Description | Default Level |
| --- | --- | --- |
| ArrayEquals | Checks for comparison of arrays using == which will always return false | Info |
| ArraysToString | Checks for explicit toString calls on arrays | Warning |
| ArraysInFormat | Checks for arrays passed to String.format | Error |

*A. Memory leaks*

One notable type of defect that is not detected by most of these tools is that of the nature of memory safety. The Scala programming language offers a single memory management model for heap-allocated memory: fully automatic garbage collection. In theory, this approach lets the developer completely forget about memory management as GC will automatically deallocate objects for them whenever it considers them to be unreachable. Unfortunately, in practice GC has non-trivial performance trade-offs that might not be acceptable in some applications [16]. Also, if the references to unused objects are present in the running process, these cannot be garbage-collected. This can trigger a memory leak. In other words, a memory leak can still occur in a memory managed language like Scala and occurs when a process maintains unnecessary references to some unused objects [17]. Detecting defects of this nature via pure syntactic analysis and pattern matching would make it extremely difficult, if not impossible, to detect in Scala. Hence detecting such defects would entail some form of semantic information, not just syntactic pattern matching. Hence for a language like Scala, a well-rounded Static analyzer that can detect a certain range of errors needs to incorporate Semantic analysis in addition to syntactic analysis.

*B. Unhandled exceptions*

Exception handling facilities in programming languages allow the programmer to define, raise and handle exceptional conditions. The exception facilities, however, can provide a hole for the program safety[18]. A program can terminate abnormally when an exception is raised and never handled [19]. Almost all mainstream programming languages provide some form of exception handling, as does Scala [20]. There has been a long discussion between language creators on whether or not exceptions should be subject to static checking [21]. While some languages do offer static checking for exceptions, which ensures that exceptions in the given source code are handled correctly (an example being Java), Scala does not provide static checking for exception handling [22]. This would mean that a Scala program with unhandled exceptions could be compiled successfully, but when unhandled exceptions do occur, this could lead to abnormal termination of the program and disrupt the normal operations. Hence detecting these unhandled exception cases during the development phase can help in creating more robust applications. As stated before, the designers of Scala opted not to include concept of checked exceptions owing mostly to the Functional programming paradigm of the language [23]. Of the available static analysis tools surveyed, Scapegoat provided the best detection of defects related to Exception handling. Scapegoat provides an inspection rule that finds catch blocks that don't handle caught exceptions and raises a warning. It also provides 4 Inspection rules related to bad programmer practices of catching certain types of exceptions: checks for try blocks that catch Exception, checks for try blocks that catch fatal exceptions (VirtualMachineError, ThreadDeath, InterruptedException, LinkageError, ControlThrowable), checks for try blocks that catch Throwable, which are all tagged as warnings and checks for try blocks that catch null pointer exceptions, which is tagged as an error. Scapegoat also has a rule that looks for empty try blocks and checks for catch clauses that cannot be reached. As SonarScala does provide configuration options to import rule sets from ScalaStyle and Scapegoat, the above-mentioned exception handling checks can also be performed using SonarScala. ScalaStyle and CheckMarx does not provide any unhandled exception checks [14].

*C. Null references*

Null references, termed the "billion-dollar mistake" by Sir Anthony Hoare [24] are widely regarded in the programming community as something that should be avoided in production code. Null is a special value that denotes a reference that points to nowhere, hence inhabiting all reference types in Scala and Java. A programmer usually uses the null reference to dereference some object, but using such techniques does not align with best coding practices. This is because there is no concrete way to establish that a reference is null until one explicitly checks it and no operation can safely be done until the check is complete, which raises an exception related to Java's NullPointerException in most programming languages. To mitigate this, Scala provides programmers with the Options class, which acts as a container that allows some value or no-value [25]. But Scala also has the Null class within its standard core libraries [26], so programmers new to Scala or unfamiliar with coding best practices could potentially use the Null class to create Null references. Of the tools surveyed, Scapegoat provided built-in inspection rules for checking the use of NULL assignments and method invocations. Scapegoat also has a rule that checks for try blocks that catch null pointer exceptions [14]. SonarScala can import the rules from Scapegoat and hence can also detect null references. WartRemover also provides built in checks that can detect null references [13]. With the documentation that is available, CheckMarx CxSAST does not provide any rules to detect Null References.

*D. Cross-site scripting*

Cross-Site-Scripting (XSS) is one of the topmost threats to web applications. It accounts for nearly 50% of all web application vulnerabilities detected since 2012 and is expected to increase by huge margins in the coming years [27]. This vulnerability type is also frequently included in the OWASP Top 10 list and in 2017 it was present in almost two thirds of all web applications [28]. There are 3 broad types of XSS attacks: Reflected XSS, Stored XSS and DOM XSS [28]. An XSS attack usually occurs when actors insert malicious code into webpages via input fields, such that proper input validations are required to overcome attacks of this nature. Performing a static analysis on the source code before deployment of the application is one of the most commonly used practices, proposed by numerous studies [29]. The survey by the authors of [29] shows that static analysis as a solution to XSS is proposed by almost 24% of the papers surveyed. Most static XSS analysis techniques use static taint analysis, a technique which tracts tainted values through a CFG. Of all the tools that were surveyed, only CheckMarx provided the capabilities to detect both Stored and Reflected XSS defects. Most research combined static taint analysis with other techniques, such as symbolic code execution [30], data-flow analysis [31] and string analysis [32], among others.

*E. XML external entity*

XML External Entity (XXE) attacks can occur when an XML parser supports XML entities while processing XML received from an untrusted source, which can then lead to various attacks, such as disclosure of confidential data, denial of service and diversion from intended purpose [33]. This type of vulnerability occurs in Scala mostly when consuming RESTful Web services. Security firm snyk in 2016 disclosed this vulnerability being present in io.spray, which is a suite of lightweight Scala libraries for building and consuming RESTful web services on top of the Akka framework [34]. The research from [35] denotes the need for a secure XML validator for functional programming languages like Scala. The authors of [34] mention that they designed a new XML input validator for the OCaml functional programming language, which can detect instances of XEE attacks. Of the tools currently available on the market, none offer detection capabilities of defects of this nature. CheckMarx CxSAST does state, however, that defects of this nature could be detected, but does not provide details on whether this is possible with their Scala plugin. Hence there is a clear lack in static detectors and input validators related to XEE vulnerability for programs written in Scala. Thus, a potential research direction would be to develop a technique using the semantic and syntactic rules of the Scala functional programming language.

*F. Multithreaded synchronization*

Since its inception, Scala was designed to run on top of JVM, which has rich and well-defined functions and libraries for multi-threaded execution. As a result, the memory model in Scala, the multi-threading capabilities and its thread synchronization capabilities are inherited from JVM [36]. Based on this fact, there are multiple inter-thread synchronization mechanisms built within Scala, the most commonly used being the synchronized statement. Ferrara et. al [37] proposes a generic static analyzer called Checkmate designed for multithreaded Java programs. It is based on the abstract interpretation theory and is tuned at the bytecode level so that it can analyze external libraries as well as programs written in languages like Scala that compile to Java bytecode. Based on initial research, Checkmate is the only available static analyzer for multithreaded programs that could handle Scala code. But as this tool relies on static analysis on the bytecode, this would imply that compiling the Scala code to bytecode is a prerequisite to using this tool, which would then defeat the purpose of a true static code analyzer which can detect defects as the programmer writes the code. Hence another potential research direction would be to design an approach that can natively handle concurrent program execution within Scala and not rely on compilation to bytecode.

Research from [1, 38, 39] has shown that in all cases one single static analysis tool cannot detect all types of severe defects present in a code base. It is to be noted that they performed their research not specifically targeting Scala, but rather more prevalent languages like C/C++ and Java. Most of the available Static Analyzers obtain different results for different types of vulnerabilities and cover different subsets of these vulnerabilities. In most cases running multiple tools leads to a larger coverage area and increases the number of errors detected. From this paper it is quite clear that there is a lack of total coverage for programs written in Scala, which provides obvious grounds for future research to develop techniques and tools that can provide better vulnerability and defect detection and reporting.

ACKNOWLEDGMENTS

This work is funded by Towards a Security Framework for the Future Heterogeneous Internet of Things: A Pilot Study, College of Technology, University of Houston, U.S.A.